%% file: main.tex
\documentclass[10pt,twocolumn,letterpaper]{article}

\usepackage[pagenumbers]{cvpr} % To force page numbers, e.g. for an arXiv version


\definecolor{cvprblue}{rgb}{0.21,0.49,0.74}
\usepackage[pagebackref,breaklinks,colorlinks,allcolors=cvprblue]{hyperref}
\usepackage{algorithm}
\usepackage{algorithmic}
\usepackage{amsmath}
\usepackage{amssymb}
\usepackage{multicol}
\usepackage{multirow}
\usepackage{makecell}
\usepackage{threeparttable}
\def\paperID{2720} % *** Enter the Paper ID here
\def\confName{CVPR}
\def\confYear{2025}

\title{Balanced Rate-Distortion Optimization in Learned Image Compression}

\author{
Yichi Zhang, Zhihao Duan, Yuning Huang, Fengqing Zhu\\
Elmore Family School of Electrical and Computer Engineering,\\
Purdue University, West Lafayette, Indiana, U.S.A.\\
{\tt\small \{zhan5096, duan90, huan1781, zhu0\}@Purdue.edu}
}

\begin{document}
\maketitle
\input{sec/0_abstract}    
\input{sec/1_intro}

\input{sec/2_related_work}

\input{sec/3_proposed_method}

\input{sec/4_experiment}

\input{sec/5_conclusion}
\clearpage
{
    \small
    \bibliographystyle{ieeenat_fullname}
    \bibliography{main}
}

% WARNING: do not forget to delete the supplementary pages from your submission 
% \input{sec/X_suppl}

\end{document}

%% file: sec/0_abstract.tex
\begin{abstract}
Learned image compression (LIC) using deep learning architectures has seen significant advancements, yet standard rate-distortion (R-D) optimization often encounters imbalanced updates due to diverse gradients of the rate and distortion objectives. This imbalance can lead to suboptimal optimization, where one objective dominates, thereby reducing overall compression efficiency. To address this challenge, we reformulate R-D optimization as a multi-objective optimization (MOO) problem and introduce two balanced R-D optimization strategies that adaptively adjust gradient updates to achieve more equitable improvements in both rate and distortion. The first proposed strategy utilizes a coarse-to-fine gradient descent approach along standard R-D optimization trajectories, making it particularly suitable for training LIC models from scratch. The second proposed strategy analytically addresses the reformulated optimization as a quadratic programming problem with an equality constraint, which is ideal for fine-tuning existing models. Experimental results demonstrate that both proposed methods enhance the R-D performance of LIC models, achieving around a 2\% BD-Rate reduction with acceptable additional training cost, leading to a more balanced and efficient optimization process. Code will be available at~\url{https://gitlab.com/viper-purdue/Balanced-RD}.
\end{abstract}

%% file: sec/1_intro.tex
\section{Introduction}
\label{sec:intro}

Rate-distortion (R-D) optimization~\cite{sullivan1998rate} is a fundamental process in image compression, balancing the trade-off between rate (compression efficiency) and distortion (image quality). Traditional compression techniques, such as JPEG~\cite{wallace1992jpeg} and MPEG~\cite{le1991mpeg}, achieve this balance using carefully crafted transforms, quantization, and entropy coding schemes. These methods reduce data size while minimizing visual degradation through predefined, hand-tuned parameters and well-established signal processing techniques.

These codecs approach R-D optimization within a structured, deterministic parameter space. Parameters are either manually adjusted or selected via straightforward heuristics rather than through complex, gradient-based updates. This setup avoids the potential for imbalanced optimization arising from dynamic parameter adjustments, ensuring a stable alignment between rate and distortion objectives~\cite{sullivan2012overview}. However, the fixed nature of these parameters limits the flexibility of traditional methods in adapting to varying image content, constraining their overall efficiency.

In contrast, learned image compression (LIC) models utilize deep learning architectures to automatically adapt to image data. By learning intricate features directly from data, LIC models achieve higher compression ratios and enhanced reconstruction quality~\cite{he2022elic,liu2023learned,li2024frequencyaware}. In LIC, R-D optimization is typically achieved by minimizing a combined loss function for rate and distortion, controlled by a weighting factor, $\lambda$, which denotes the trade-off between them~\cite{ballé2018variational,balle2020nonlinear}. This gradient-based approach dynamically updates parameters to minimize the loss, enabling LIC models to adapt to diverse image content. However, summing the rate and distortion gradients in this setup can result in imbalanced updates, where one objective dominates the other due to the diverse gradient direction or magnitude during the optimization process~\cite{alemi2018fixing,sener2018multi}. This imbalance may hinder convergence, as one objective is prioritized at the expense of the overall performance drop~\cite{yu2020gradient}.

\begin{figure*}[t]
    \centering
    \begin{subfigure}[b]{0.49\textwidth}
        \centering
        \includegraphics[width=\textwidth]{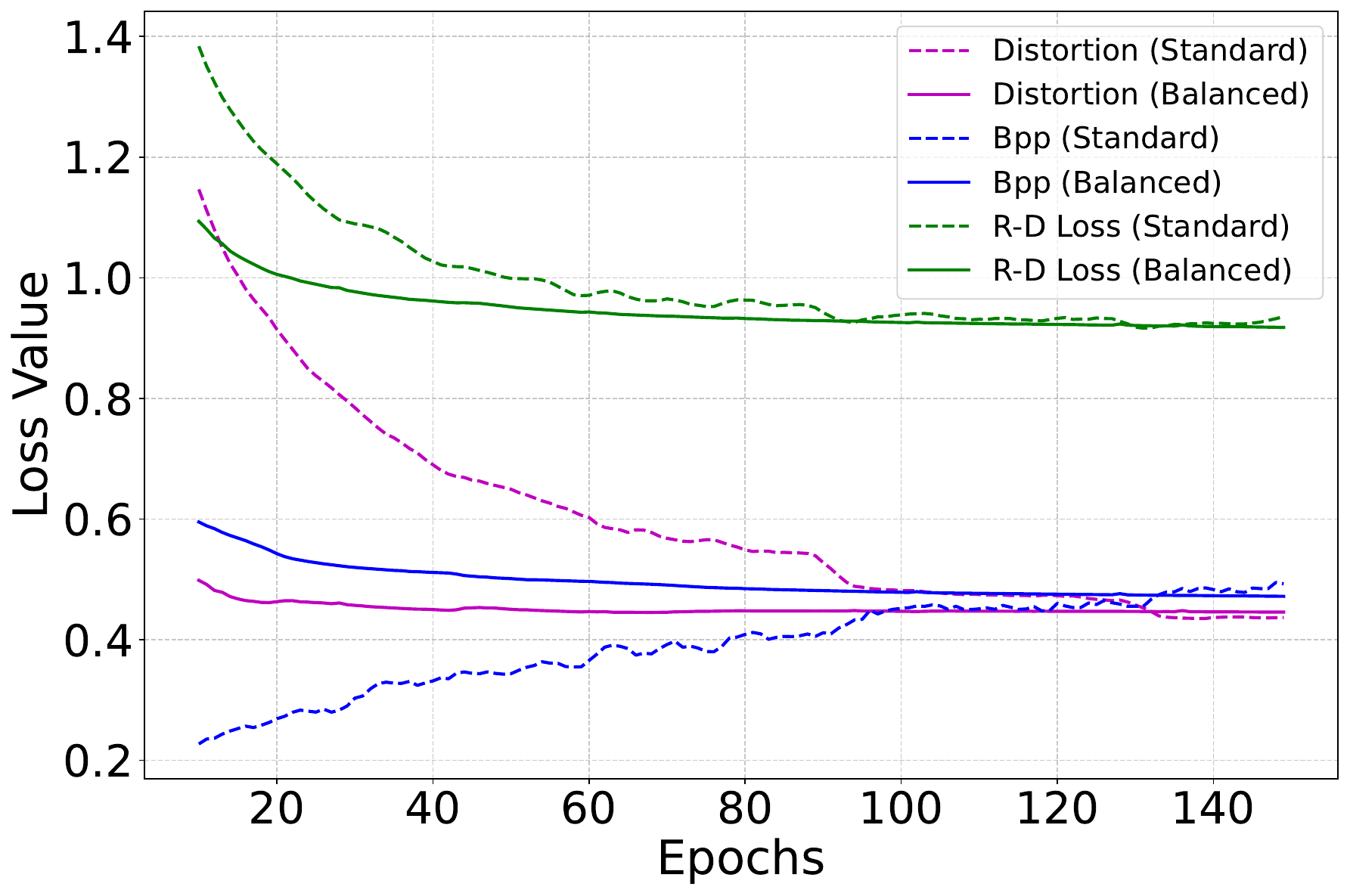}
        \caption{Testing loss trend}
        \label{subfig:std}
    \end{subfigure}
    \hfill
    \begin{subfigure}[b]{0.49\textwidth}
        \centering
        \includegraphics[width=\textwidth]{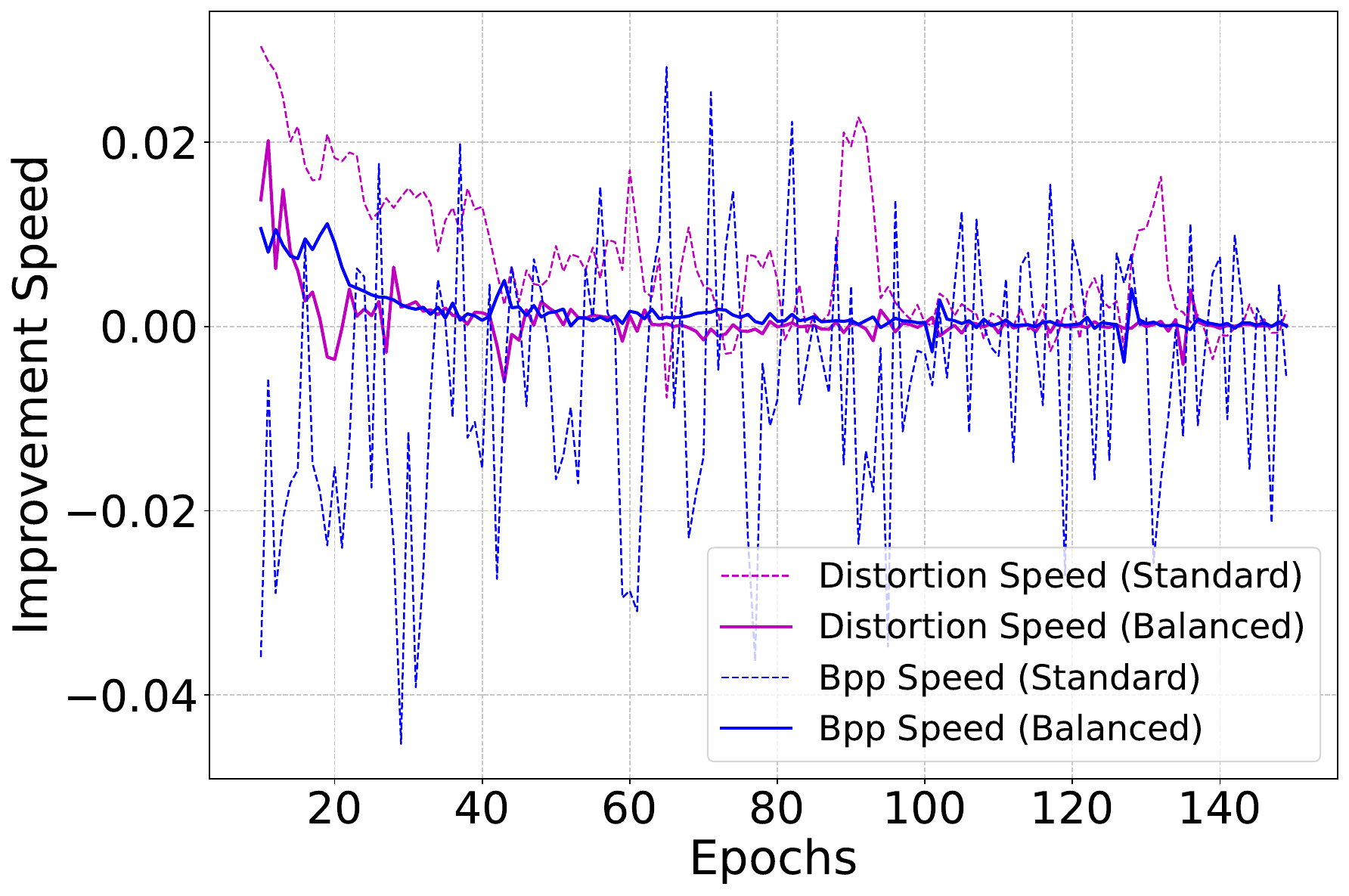}
        \caption{Loss improvement speed}
        \label{subfig:bal}
    \end{subfigure}
\caption{Comparison of loss trends and improvement speeds for the M\&S hyperprior~\cite{minnen2018joint} model with $\lambda = 0.013$ from epoch 10 to 150, where Distortion = MSE $\times \lambda \times 255^2$. The first 10 epochs are omitted for improved readability. (a) Testing loss trend for the standard R-D optimization versus the proposed balanced R-D optimization. The balanced approach demonstrates more stable and smoother loss improvements, showing simultaneous and consistent reductions in both distortion and bits-per-pixel (bpp) metrics. In contrast, the standard method focuses on reducing distortion, with the bpp loss only gradually increasing in this period. (b) Comparison of loss improvement speeds, Eq.~\ref{eq:speed}, for standard and balanced R-D optimizations. The balanced R-D optimization yields a more consistent, less volatile improvement speed, outperforming the standard approach by achieving steady and reliable convergence across epochs. This highlights the advantage of the balanced approach in optimizing both objectives cohesively and effectively.}
    \label{fig:R_D_O}
\end{figure*}

Addressing this imbalance requires strategies to harmonize rate and distortion objectives, enabling both to be optimized effectively for higher-quality image compression. Multi-objective optimization (MOO) strategies offer promising solutions, as they are designed to make progress across multiple objectives simultaneously~\cite{desideri2012multiple,sener2018multi,yu2020gradient,navon22a,liu2024famo}. These approaches guide models towards Pareto-optimal solutions~\cite{miettinen1999nonlinear} that balance competing objectives without excessively prioritizing any single one.

In this work, we introduce a MOO-based framework for R-D optimization in LIC, treating rate and distortion as two objectives to be optimized concurrently. We propose two balanced solutions for this framework that adaptively re-weight gradient updates from the two objectives to ensure equitable progress on both objectives. The first solution employs a coarse-to-fine gradient descent approach, particularly effective for training models from scratch by iteratively refining gradient weights. The second solution reformulates the optimization as a equality constraint-quadratic programming problem to derive the gradient weights analytically, offering a precise fine-tuning mechanism for existing LIC models. Our approach achieves balanced progress in both objectives, yielding smoother convergence and enhanced R-D performance, as illustrated in Fig.~\ref{fig:R_D_O}.

Our main contributions are summarized as follows: \begin{itemize} 
\item We reformulate R-D optimization as a multi-objective optimization (MOO) problem, introducing a balanced framework to optimize rate and distortion more equitably. 

% \item We propose two balanced solutions: a coarse-to-fine gradient descent solution for new model training, and an analytical quadratic programming solution for fine-tuning existing models. 
\item {We propose two solutions for balanced R-D optimization: a coarse-to-fine gradient descent solution for training new models from scratch, and an analytical quadratic programming solution for fine-tuning existing models.}

% \item Extensive experiments validate the effectiveness of our approach in improving R-D performance in LIC models. 
\item {Extensive experiments show that our approach consistently improves the R-D performance of LIC models.} 
\end{itemize}

%% file: sec/2_related_work.tex
\section{Related work}
\label{sec:related}
\subsection{Learned image compression}
Learned image compression methods often rely on non-linear transform coding framework~\cite{balle2020nonlinear} to optimize the trade-off between bit-rate R and distortion D. Recent advances in LIC have focused on two main aspects:

\begin{itemize}
    \item \textbf{Transform Functions}: Various architectures and techniques are utilized to enhance the capacity of the transform functions. For instance, residual networks~\cite{he2022elic, cheng2020learned}, deformable convolutions~\cite{fu2024fast,zhang2024efficient}, and frequency decomposition~\cite{fu2024weconvene, ma2020end}. Invertible neural networks are explored in~\cite{xie2021enhanced, cai2024i2c}, while contextual clustering is leveraged in~\cite{zhang2024another, qi2024long}. Additionally, transformers are increasingly incorporated for further performance gains~\cite{liu2023learned, zhu2022transformer, zou2022devil, koyuncu2022contextformer, qian2022entroformer, li2024frequencyaware}.
    
    \item \textbf{Entropy Model Refinement}: Improvements in entropy models have involved hierarchical priors~\cite{ballé2018variational, hu2020coarse, duan2023qarv}, spatial autoregressive models~\cite{minnen2018joint}, channel autoregressive models~\cite{minnen2020channel}, and joint channel-spatial context models~\cite{jiang2023mlic, ma2021cross}. Methods like the two-pass checkerboard approach~\cite{he2021checkerboard}, codebooks~\cite{zhu2022unified}, and (lattice) vector quantization~\cite{ zhang2023lvqac, feng2023nvtc} are also explored to enhance efficiency.
\end{itemize}

Efforts have also focused on efficient implementation in LICs, such as slimmable sub-CAEs~\cite{Tao_2023_ICCV}, variable-bit-rate codecs~\cite{Wang2023evc, kamisli2024dcc_vbrlic}, knowledge distillation~\cite{fu2024fast,zhang2024theoretical}, shallow and linear decoders~\cite{yang2023computationally}, causal context losses~\cite{han2024causal}, and latent decorrelation losses~\cite{ALi2023towards}. In addition, studies on rate-distortion-complexity~\cite{minnen2023advancing, gao2024exploring, zhang2024efficient} seek to optimize the trade-off between computational cost and rate-distortion performance. Despite these advancements in BD-Rate reduction, a thorough exploration of R-D optimization in LICs is still needed.

\subsection{Multi-objective optimization}

Multi-objective optimization aims to minimize multiple objectives with a single solution. In non-trivial problems, there is typically no single solution that can optimize all objective functions simultaneously~\cite{miettinen1999nonlinear}. Numerous optimization methods have been proposed to approximate the Pareto set and front by generating a finite set of solutions~\cite{zhou2011multiobjective, li2015many}.

When all objective functions are differentiable, the gradient-based approach can be used to find a valid gradient direction that allows simultaneous improvement across all objectives~\cite{fliege2000steepest, schaffler2002stochastic, desideri2012mutiple} according to multi-objective Karush–Kuhn–Tucker
(KKT) conditions~\cite{mangasarian1994nonlinear}. The Multiple Gradient Descent Algorithm (MGDA)~\cite{desideri2012mutiple, sener2018multi} achieves this by calculating a valid gradient \( d_t = \sum_{i=1} w_i \nabla f_i(x) \) through solving the following quadratic programming problem at each iteration:
\begin{equation}
    \min_{w_i} \left\| \sum_{i=1} w_i \nabla f_i(x_t) \right\|_2^2, \quad \text{s.t.} \quad \sum_{i=1} w_i = 1, \quad w_i \geq 0.
\end{equation}

The current solution is then updated by gradient descent as \( \theta_{t+1} = \theta_t - \eta_t d_t \). If \( d_t = 0 \), it implies there is no valid gradient direction that can improve all objectives simultaneously, making \( \theta_t \) a Pareto stationary solution~\cite{desideri2012mutiple, fliege2019complexity}. This approach has also inspired various adaptive gradient methods in multi-task learning~\cite{yu2020gradient, liu2021conflict, momma2022multi, navon22a, zhou2022convergence, senushkin2023independent, fernando2023mitigating, liu2024famo, chen2024three, xiao2024direction, hu2024revisiting}.
These works highlight the promising performance of multi-objective optimization, which has not yet been thoroughly explored in the context of learned image compression R-D optimization.

%% file: sec/3_proposed_method.tex
\section{Proposed method}
% \subsection{Preliminaries on LIC rate-distortion optimization}
\subsection{Preliminaries: rate-distortion optimization in learned image compression}

In LIC, the objective is to encode an image \( x \), drawn from a source distribution with probability density function \( p_{\text{source}} \), into a compact bit sequence for efficient storage or transmission. The receiver then reconstructs an approximation \( \hat{x} \) of the original image \( x \). The LIC process involves three main steps: encoding and quantization, entropy coding, and reconstruction.

\begin{itemize}
    \item \textbf{Encoding and Quantization}: First, each data point \( x \) is mapped to a de-correlated low-dimensional latent variable \( \hat{z} \) via an encoder function \( e(\cdot) \) followed by quantization \( Q(\cdot) \) to convert the continuous representation into discrete values, i.e., \( \hat{z} = Q(e(x)) \).

    \item \textbf{Entropy Coding}: After determining \( \hat{z} \), lossless entropy coding, such as Huffman coding~\cite{moffat2019huffman} or arithmetic coding~\cite{witten1987arithmetic}, is applied to produce a compressed bit sequence with length \( b(\hat{z}) \). Ideally, the entropy coding scheme approximates the theoretical bit rate, given by the entropy of \( \hat{z} \) under its marginal distribution. We assume that both the sender and receiver have access to an entropy model \( P(\hat{z}) \), which estimates the marginal probability distribution of \( \hat{z} \) and determines the expected bit length as \( b(\hat{z}) \approx -\log_2 P(\hat{z}) \). The goal is to ensure that \( P(\hat{z}) \) closely approximates the true marginal distribution \( p(\hat{z}) \), which is defined as
\(
p(\hat{z}) = \mathbb{E}_{x \sim p_{\text{source}}} \left[ \delta(\hat{z}, Q(e(x))) \right],
\)
where \( \delta \) is the Kronecker delta function. Under this approximation, the encoding length \( b(\hat{z}) \) is nearly optimal, as the average code length approaches the entropy \( -\log_2 p(\hat{z}) \) of \( \hat{z} \).

    \item \textbf{Reconstruction}: Once the receiver has obtained \( \hat{z} \), it reconstructs the approximation \( \hat{x} \) using a reconstruction function \( r(\cdot) \), such that \( \hat{x} = r(\hat{z}) \).
\end{itemize}

To optimize the LIC scheme, our goal is to minimize both the bit rate (rate) and the discrepancy between \( x \) and \( \hat{x} \) (distortion), where the distortion is measured by a function \( d(x, \hat{x}) \). This objective is formulated as a R-D loss using a Lagrangian multiplier:
\begin{equation}
\label{eq:rd}
    \mathcal{L}_\text{R-D} = \mathbb{E}_{x \sim p_{\text{source}}} \left[ \underbrace{-\log_2 P(\hat{z})}_{\text{rate}\ (\mathcal{L}_{\text{R}})} + \underbrace{\lambda d(x, \hat{x})}_{\text{distortion}\ (\mathcal{L}_{\text{D}})} \right],
\end{equation}
where \( \lambda \) is a Lagrange multiplier that controls the trade-off between rate (compression efficiency) and distortion (reconstruction quality). This formulation represents the Lagrangian relaxation of the distortion-constrained R-D optimization problem, aiming for efficient compression while maintaining high reconstruction fidelity. In practice, the gradient used to update LIC models is the sum of the gradients for each objective:
\(
d_t = \nabla \mathcal{L}_{\text{R}, t} + \nabla \mathcal{L}_{\text{D}, t}
\), where $t$ denotes the training iteration.

\begin{algorithm*}[t]
\caption{Balanced Rate-Distortion Optimization via Trajectory Optimization}
\label{alg:solution1}
\begin{algorithmic}[1]
\REQUIRE Initial network parameters $\theta_0$, initial softmax logits $\boldsymbol{\xi}_0$, learning rates $\alpha$ (for $\theta$) and $\beta$ (for $\boldsymbol{\xi}$), decay parameter $\gamma$, total iterations $T$
\STATE Initialize weights $\mathbf{w}_0 \leftarrow \text{Softmax}(\boldsymbol{\xi}_0)$
\FOR{$t = 0$ \TO $T-1$}
    \STATE Compute losses $\mathcal{L}_{\text{R}, t} = \mathcal{L}_{\text{R}}(\theta_t)$ and $\mathcal{L}_{\text{D}, t} = \mathcal{L}_{\text{D}}(\theta_t)$
    \STATE Compute gradients: $\nabla_\theta \mathcal{L}_{\text{R}, t} = \frac{\partial \mathcal{L}_{\text{R}, t}}{\partial \theta_t}$, \quad $\nabla_\theta \mathcal{L}_{\text{D}, t} = \frac{\partial \mathcal{L}_{\text{D}, t}}{\partial \theta_t}$
    \STATE \hspace{2.1cm} $\nabla_\theta \log \mathcal{L}_{\text{R}, t} = \frac{\nabla_\theta \mathcal{L}_{\text{R}, t}}{\mathcal{L}_{\text{R}, t}}$, \quad $\nabla_\theta \log \mathcal{L}_{\text{D}, t} = \frac{\nabla_\theta \mathcal{L}_{\text{D}, t}}{\mathcal{L}_{\text{D}, t}}$
    \STATE Compute normalization constant: $c_t = \left( \frac{w_{\text{R}, t}}{\mathcal{L}_{\text{R}, t}} + \frac{w_{\text{D}, t}}{\mathcal{L}_{\text{D}, t}} \right)^{-1}$
    \STATE Compute balanced gradient: $\mathbf{d}_t = c_t \left( w_{\text{R}, t} \nabla_\theta \log \mathcal{L}_{\text{R}, t} + w_{\text{D}, t} \nabla_\theta \log \mathcal{L}_{\text{D}, t} \right)$
    \STATE Update network parameters: $\theta_{t+1} = \theta_t - \alpha\, \mathbf{d}_t$
    \STATE Compute updated losses $\mathcal{L}_{\text{R}, t+1} = \mathcal{L}_{\text{R}}(\theta_{t+1})$ and $\mathcal{L}_{\text{D}, t+1} = \mathcal{L}_{\text{D}}(\theta_{t+1})$
    \STATE Compute $\delta_t = \begin{bmatrix} \frac{\partial w_{\text{R}, t}}{\partial \boldsymbol{\xi}_t} \\ \frac{\partial w_{\text{D}, t}}{\partial \boldsymbol{\xi}_t} \end{bmatrix}^\top \begin{bmatrix} \log \mathcal{L}_{\text{R}, t} - \log \mathcal{L}_{\text{R}, t+1} \\ \log \mathcal{L}_{\text{D}, t} - \log \mathcal{L}_{\text{D}, t+1} \end{bmatrix}$
    \STATE Update softmax logits: $\boldsymbol{\xi}_{t+1} = \boldsymbol{\xi}_t - \beta \left( \delta_t + \gamma\, \boldsymbol{\xi}_t \right)$
    \STATE Update weights: $\mathbf{w}_{t+1} = \text{Softmax}(\boldsymbol{\xi}_{t+1})$
\ENDFOR
\end{algorithmic}
\end{algorithm*}

\subsection{Balanced rate-distortion optimization}
To address the imbalance in optimizing rate and distortion discussed in Sec.~\ref{sec:intro}, we propose a balanced R-D optimization framework. This framework dynamically adjusts the contributions from each objective in the R-D loss function, aiming to achieve equal progress in both rate and distortion. Specifically, we redefine the update direction as \( d_t = w_{\text{R},t} \nabla \log \mathcal{L}_{\text{R}, t} + w_{\text{D},t} \nabla \log \mathcal{L}_{\text{D}, t}\), where \(w_{\text{R},t}\) and \(w_{\text{L},t}\) are adaptive weights for the rate and distortion gradients. By maximizing the minimum improvement speed between rate and distortion at each step, our method finds the ideal gradient weights, promotes stable balanced optimization, and enhances overall model performance.

Inspired by  MOO techniques~\cite{sener2018multi, liu2024famo}, we reformulate the R-D optimization problem as a multi-objective optimization problem. Our goal is to simultaneously optimize rate minimization and distortion minimization with parameters \( \theta \in \mathbb{R}^m \). To achieve this, we modify the original Lagrangian R-D loss function to the following form:
\begin{equation}
\min_{\theta \in \mathbb{R}^m} \left\{ \mathcal{L}(\theta) = \mathcal{L}_{\text{R}}(\theta) + \mathcal{L}_{\text{D}}(\theta)  \right\},
\end{equation}
where the trade-off factor \( \lambda \) is now absorbed into \( \mathcal{L}_{\text{D}} \) to simplify the formulation. $m$ is the number of parameters.

The loss improvement speed at iteration \( t \) of task \(i \in \{\text{R}, \text{D}\}\) is defined as:
\begin{equation}
\label{eq:speed}
s_{i,t}(\alpha, d_t) = \frac{\mathcal{L}_{i,t} - \mathcal{L}_{i,t+1}}{\mathcal{L}_{i,t}},
\end{equation}
where \( \alpha \) is the step size and \( d_t \) is the update direction at \( t \), with \( \theta_{t+1} = \theta_t - \alpha d_t \). To achieve balanced optimization, we seek an update direction \( d_t \) that maximizes the minimum improvement speed between rate and distortion, ensuring that neither objective dominates the update. In other words, we maximize the smaller improvement speed between \( \mathcal{L}_{\text{R}} \) and \( \mathcal{L}_{\text{D}} \). This leads to the following saddle point problem~\cite{lin2020near} formulation:
{\small
\begin{equation}
\max_{d_t \in \mathbb{R}^m} \min \left( \frac{1}{\alpha} s_{\text{R},t}(\alpha, d_t) - \frac{1}{2} \|d_t\|^2, \frac{1}{\alpha} s_{\text{D},t}(\alpha, d_t) - \frac{1}{2} \|d_t\|^2 \right),
\end{equation}}
where \( \frac{1}{2} \|d_t\|^2 \) serves as a regularization term to prevent unbounded updates.
% where \( \frac{1}{2} \|d_t\|^2 \) serves as a regularization term to prevent an unbounded solution.

When the step size \( \alpha \) is small, one can approximate \( \mathcal{L}_{t+1} \approx \mathcal{L}_{t} - \alpha \nabla \mathcal{L}_t^\top d_{t} \) using a first-order Taylor expansion. This approximation simplifies the problem to:
{\small\begin{equation}
\begin{aligned}
&\max_{d_t \in \mathbb{R}^m} \min \left( \frac{1}{\alpha} s_{\text{R},t}(\alpha, d_t) - \frac{1}{2} \|d_t\|^2, \frac{1}{\alpha} s_{\text{D},t}(\alpha, d_t) - \frac{1}{2} \|d_t\|^2 \right)\\
&=\max_{d_t \in \mathbb{R}^m} \min \left( \frac{\nabla \mathcal{L}_{\text{R}, t}^\top d_t}{\mathcal{L}_{\text{R}, t}} - \frac{1}{2} \|d_t\|^2, \frac{\nabla \mathcal{L}_{\text{D}, t}^\top d_t}{\mathcal{L}_{\text{D}, t}} - \frac{1}{2} \|d_t\|^2 \right) \\
&= \max_{d_t \in \mathbb{R}^m} \left( \min \left( \nabla \log \mathcal{L}_{\text{R}, t}^\top d_t, \nabla \log \mathcal{L}_{\text{D}, t}^\top d_t \right) - \frac{1}{2} \|d_t\|^2 \right).
\end{aligned}
\end{equation}}

To avoid solving the high-dimensional primal problem directly (as \( d_t \in \mathbb{R}^m \) with potentially millions of parameters if \( \theta \) is a neural network), we follow previous work~\cite{sener2018multi,liu2024famo} to turn to the dual problem. Leveraging the Lagrangian duality theorem~\cite{boyd2004convex}, we can rewrite the optimization as a convex combination of gradients:
{\small
\begin{equation}
\begin{aligned}
&\max_{d_t \in \mathbb{R}^m} \left( \min \left( \nabla \log \mathcal{L}_{\text{R}, t}^\top d_t, \nabla \log \mathcal{L}_{\text{D}, t}^\top d_t \right) - \frac{1}{2} \|d_t\|^2 \right) \\
&= \max_{d_t \in \mathbb{R}^m} \min_{w \in \mathbb{S}_2} \left( w_{\text{R},t} \nabla \log \mathcal{L}_{\text{R}, t} + w_{\text{D},t} \nabla \log \mathcal{L}_{\text{D}, t} \right)^\top d_t - \frac{1}{2} \|d_t\|^2 \\
&= \min_{w_t \in \mathbb{S}_2} \max_{d_t \in \mathbb{R}^m} \left( w_{\text{R},t} \nabla \log \mathcal{L}_{\text{R}, t} + w_{\text{D},t} \nabla \log \mathcal{L}_{\text{D}, t} \right)^\top d_t - \frac{1}{2} \|d_t\|^2,
\end{aligned}
\end{equation}}
where the second equality follows from strong duality. Here, \( w_t \in \mathbb{S}_2 = \{ w \in \mathbb{R}_{\geq 0}^2 \mid w^\top \mathbf{1} = 1 \} \) represents the gradient weights in the 2-dimensional probabilistic simplex.

Let \( g(d_t, w) = (w_{\text{R},t} \nabla \log \mathcal{L}_{\text{R}, t} + w_{\text{D},t} \nabla \log \mathcal{L}_{\text{D}, t})^\top d_t - \frac{1}{2} \|d_t\|^2 \). The optimal direction \( d_t^* \) is obtained by setting:
\begin{equation}
\small
\label{eq:opt_grad}
\frac{\partial g}{\partial d_t} = 0 \quad \Longrightarrow \quad d_t^* = w_{\text{R},t} \nabla \log \mathcal{L}_{\text{R}, t} + w_{\text{D},t} \nabla \log \mathcal{L}_{\text{D}, t}.
\end{equation}

Substituting \( d_t^* \) back, we obtain:
\begin{equation}
\small
\label{eq:target}
\begin{aligned}
&\max_{d_t \in \mathbb{R}^m} \min \left( \frac{1}{\alpha} s_{\text{R},t}(\alpha, d_t) - \frac{1}{2} \|d_t\|^2, \frac{1}{\alpha} s_{\text{D},t}(\alpha, d_t) - \frac{1}{2} \|d_t\|^2 \right) \\
&= \min_{w_t \in \mathbb{S}_2} \frac{1}{2} \left\| w_{\text{R},t} \nabla \log \mathcal{L}_{\text{R}, t} + w_{\text{D},t} \nabla \log \mathcal{L}_{\text{D}, t} \right\|^2 \\
&= \min_{w_t \in \mathbb{S}_2} \frac{1}{2} \| J_t w_t \|^2,
\end{aligned}
\end{equation}
where \( J_t = \begin{bmatrix}
\nabla \log \mathcal{L}_{\text{R}, t}^\top \\
\nabla \log \mathcal{L}_{\text{D}, t}^\top
\end{bmatrix} \). Thus, our optimization problem reduces to finding \( w_t \) that satisfies Eq.~\ref{eq:target}.

\subsubsection{Solution 1: Gradient descent over trajectory}
\label{subsubsec:gd_over}
Rather than fully solving the optimization problem at each step, our proposed Solution 1 adopts a coarse-to-fine gradient descent approach~\cite{sener2018multi,liu2024famo}, incrementally refining the solution along the R-D optimization trajectory. In this approach, the gradient weights \( w_t \) are updated iteratively as follows:
\begin{equation}
w_{t+1} = w_{t} - \alpha_w \tilde{\delta},
\end{equation}
where
\begin{equation}
\tilde{\delta} = \nabla_w \frac{1}{2} \| J_t w_t \|^2 = J_t^\top J_t w_t.
\end{equation}

Using a first-order Taylor approximation, \( \log \mathcal{L}_{t+1} \approx \log \mathcal{L}_{t} - \alpha \nabla \log \mathcal{L}_t^\top d_{t} \), we have the following relationship:
\begin{equation}
\tilde{\delta} = J_t^\top J_t w_t = J_t^\top d_t \approx \frac{1}{\alpha} \begin{bmatrix} \log \mathcal{L}_{\text{R}, t} - \log \mathcal{L}_{\text{R}, t+1} \\ \log \mathcal{L}_{\text{D}, t} - \log \mathcal{L}_{\text{D}, t+1} \end{bmatrix}.
\footnote{To prevent negative values in \( \log \mathcal{L}_{i,t} \), we add 1 to \( \mathcal{L}_{i,t} \) before applying the logarithm, ensuring that the minimum value in the log domain is zero.}
\end{equation}

To ensure that \( w_t \) remains within the interior point of simplex \( \mathbb{S}_2 \), we reparametrize \( w_t \) using \( \xi_t \):
\begin{equation}
w_t = \text{Softmax}(\xi_t),
\end{equation}
where \( \xi_t \in \mathbb{R}^2 \) represents the unconstrained softmax logits. To give more weight to recent updates, we add a decay term~\cite{zhou2022convergence,liu2024famo}, leading to the following update for \( \xi_t \):
\begin{equation}
\xi_{t+1} = \xi_t - \beta(\delta_t + \gamma \xi_t),
\end{equation}
where
\begin{equation}
\delta_t = \begin{bmatrix} \nabla^\top w_{\text{R},t}(\xi) \\ \nabla^\top w_{\text{D},t}(\xi) \end{bmatrix} \begin{bmatrix} \log \mathcal{L}_{\text{R}, t} - \log \mathcal{L}_{\text{R}, t+1} \\ \log \mathcal{L}_{\text{D}, t} - \log \mathcal{L}_{\text{D}, t+1} \end{bmatrix}.
\end{equation}

After computing the weights \( w_t \), we renormalize them to ensure numerical stability~\cite{liu2024famo}. This renormalization is crucial as our update direction is a convex combination of the gradients of the log losses:
\begin{equation}
d_t = w_{\text{R},t} \nabla \log \mathcal{L}_{\text{R},t} + w_{\text{D},t} \nabla \log \mathcal{L}_{\text{D},t} = \sum_{i \in \{\text{R}, \text{D}\}} \frac{w_{i,t}}{\mathcal{L}_{i,t}} \nabla \mathcal{L}_{i,t}.
\end{equation}
When \( \mathcal{L}_{i,t} \) becomes small, the multiplicative factor \( \frac{w_{i,t}}{\mathcal{L}_{i,t}} \) can grow large, potentially causing instability in the optimization. To mitigate this, we scale the gradient by a constant \( c_t \):
\begin{equation}
c_t = \left(\frac{w_{\text{R},t}}{\mathcal{L}_{\text{R},t}} + \frac{w_{\text{D},t}}{\mathcal{L}_{\text{D},t}}\right)^{-1}.
\end{equation}

The resulting balanced gradient, which is used to update the model parameters \( \theta \), is then given by:
\begin{equation}
\label{eq:sl1_d}
d_t = c_t \left(w_{\text{R},t} \nabla \log \mathcal{L}_{\text{R},t} + w_{\text{D},t} \nabla \log \mathcal{L}_{\text{D},t}\right).
\end{equation}

This method, referred to as Solution 1, is a coarse-to-fine gradient descent technique. It is particularly suited for training LIC models from scratch, as it incrementally balances rate and distortion optimization along the trajectory. The implementation is shown in Algorithm~\ref{alg:solution1}.

\begin{algorithm*}[t]
\caption{Balanced Rate-Distortion Optimization via Quadratic Programming}
\label{alg:solution2}
\begin{algorithmic}[1]
\REQUIRE Initial network parameters $\theta_0$, learning rate $\alpha$, total iterations $T$
\FOR{$t = 0$ \TO $T-1$}
    \STATE Compute losses $\mathcal{L}_{\text{R}, t} = \mathcal{L}_{\text{R}}(\theta_t)$ and $\mathcal{L}_{\text{D}, t} = \mathcal{L}_{\text{D}}(\theta_t)$
    \STATE Compute gradients: $\nabla_\theta \mathcal{L}_{\text{R}, t} = \frac{\partial \mathcal{L}_{\text{R}, t}}{\partial \theta_t}$, \quad $\nabla_\theta \mathcal{L}_{\text{D}, t} = \frac{\partial \mathcal{L}_{\text{D}, t}}{\partial \theta_t}$
    \STATE \hspace{2.1cm} $\nabla_\theta \log \mathcal{L}_{\text{R}, t} = \frac{\nabla_\theta \mathcal{L}_{\text{R}, t}}{\mathcal{L}_{\text{R}, t}}$, \quad $\nabla_\theta \log \mathcal{L}_{\text{D}, t} = \frac{\nabla_\theta \mathcal{L}_{\text{D}, t}}{\mathcal{L}_{\text{D}, t}}$
    \STATE Form matrix $J_t = \begin{bmatrix}  \nabla_\theta \log \mathcal{L}_{\text{R}, t} ^\top \\  \nabla_\theta \log \mathcal{L}_{\text{D}, t} ^\top \end{bmatrix}$
    \STATE Compute Hessian matrix: $Q = J_t^\top J_t = \begin{bmatrix} \| \nabla_\theta \log \mathcal{L}_{\text{R}, t} \|^2 & \langle \nabla_\theta \log \mathcal{L}_{\text{R}, t}, \nabla_\theta \log \mathcal{L}_{\text{D}, t} \rangle \\ \langle \nabla_\theta \log \mathcal{L}_{\text{R}, t}, \nabla_\theta \log \mathcal{L}_{\text{D}, t} \rangle & \| \nabla_\theta \log \mathcal{L}_{\text{D}, t} \|^2 \end{bmatrix}$
    \STATE Compute inverse $Q^{-1}$
    \STATE Compute weights: $\lambda = \frac{1}{\mathbf{1}^\top Q^{-1} \mathbf{1}}$, \quad $w_t = \lambda\, Q^{-1} \mathbf{1}$
    \STATE Apply softmax for numerical stability: $\tilde{w}_t = \text{Softmax}({w}_t)$
    \STATE Compute normalization constant: $c_t = \left( \frac{\tilde{w}_{\text{R}, t}}{\mathcal{L}_{\text{R}, t}} + \frac{\tilde{w}_{\text{D}, t}}{\mathcal{L}_{\text{D}, t}} \right)^{-1}$
    \STATE Compute balanced gradient: $\mathbf{d}_t = c_t \left( \tilde{w}_{\text{R}, t} \nabla_\theta \log \mathcal{L}_{\text{R}, t} + \tilde{w}_{\text{D}, t} \nabla_\theta \log \mathcal{L}_{\text{D}, t} \right)$
    \STATE Update network parameters: $\theta_{t+1} = \theta_t - \alpha\, \mathbf{d}_t$
\ENDFOR
\end{algorithmic}
\end{algorithm*}

\subsubsection{Solution 2: Quadratic programming}
Alternatively, we can formulate the weight optimization problem as a constrained-quadratic programming (QP) problem~\cite{nocedal1999numerical}:
\begin{equation}
\begin{aligned}
\min_{w_t} \quad & \frac{1}{2} \| J_t w_t \|^2 = \frac{1}{2} w_t^\top (J_t^\top J_t) w_t \\
\text{s.t.} \quad & w_{\text{R},t} + w_{\text{D},t} = 1, \\
& w_{\text{R},t}, w_{\text{D},t} \geq 0,
\end{aligned}
\end{equation}
where \( J_t = \begin{bmatrix} \nabla \log \mathcal{L}_{\text{R}, t}^\top \\ \nabla \log \mathcal{L}_{\text{D}, t}^\top \end{bmatrix} \) is an \( n \times 2 \) matrix containing the gradients of the log losses for rate and distortion.

Let \( Q = J_t^\top J_t \) represent the Hessian matrix. Since the gradients for rate and distortion typically reflect distinct objectives, they are rarely parallel, suggesting \( Q \) is positive definite~\cite{sener2018multi}. This property permits an analytical solution to the QP problem. The Hessian \( Q \) is given by:
\begin{equation}
Q = \begin{bmatrix} \|\nabla \log \mathcal{L}_{\text{R}, t}\|^2 & \langle \nabla \log \mathcal{L}_{\text{R}, t}, \nabla \log \mathcal{L}_{\text{D}, t} \rangle \\ \langle \nabla \log \mathcal{L}_{\text{R}, t}, \nabla \log \mathcal{L}_{\text{D}, t} \rangle & \|\nabla \log \mathcal{L}_{\text{D}, t}\|^2 \end{bmatrix}.
\end{equation}

% When the gradients are not parallel, the determinant of \( Q \) is positive, confirming that \( Q \) is positive definite.
To solve this QP problem, we introduce a Lagrange multiplier \( \lambda \) to enforce the equality constraint~\cite{boyd2004convex}, giving us the following Lagrangian:
\begin{equation}
L(w_t, \lambda) = \frac{1}{2} w_t^\top Q w_t - \lambda(w_{\text{R},t} + w_{\text{D},t} - 1).
\end{equation}

The weights \( w_t \) can then be obtained by applying the Karush-Kuhn-Tucker (KKT) conditions~\cite{mangasarian1994nonlinear}. The main KKT conditions for this problem are:

1. \textbf{Stationarity}: The gradient of the Lagrangian with respect to \( w_t \) must be zero,
   \begin{equation}
   \nabla_{w_t} L = Q w_t - \lambda \mathbf{1} = 0,
   \end{equation}
   where \( \mathbf{1} = [1, 1]^\top \). This yields \( w_t = \lambda Q^{-1} \mathbf{1} \).

2. \textbf{Primal Feasibility}: The weights must satisfy the equality constraint,
   \begin{equation}
   \mathbf{1}^\top w_t = 1.
   \end{equation}
Note: Neglecting dual feasibility and complementary slackness simplifies the solution process since non-negativity is guaranteed by a softmax projection applied subsequently.

By combining these conditions, we arrive at a simplified expression:
\begin{equation}
\mathbf{1}^\top w_t = \lambda \mathbf{1}^\top Q^{-1} \mathbf{1} = 1,
\end{equation}
which gives:
\begin{equation}
\lambda = \frac{1}{\mathbf{1}^\top Q^{-1} \mathbf{1}}, \quad w_t = \frac{Q^{-1} \mathbf{1}}{\mathbf{1}^\top Q^{-1} \mathbf{1}}.
\end{equation}

Since \( Q \) is positive definite, \( Q^{-1} \) exists, ensuring this solution is unique. To enforce non-negativity and enhance numerical stability, we then further project \( w_t \) onto the probability simplex \( \mathbb{S}_2 \) by simply using the softmax function:
\begin{equation}
\tilde{w}_t = \text{Softmax}\left({w}_t\right)=\text{Softmax}\left(\frac{Q^{-1} \mathbf{1}}{\mathbf{1}^\top Q^{-1} \mathbf{1}}\right).
\end{equation}
This step ensures \( \tilde{w}_t \in \mathbb{S}_2 \) and reduces potential numerical issues caused by large gradient variations.  It serves as an approximation that aligns with the optimal solution while maintaining similar performance despite slight deviations.

After determining the weights \( \tilde{w}_t \), we apply the same renormalization constant \( c_t \) as in Section~\ref{subsubsec:gd_over} to compute a balanced gradient for updating the model parameters \( \theta \), following the form in Eq.~\ref{eq:sl1_d}. Solution 2, with its analytical QP formulation, is particularly suitable for fine-tuning existing LIC models, offering a refined approach to balance rate and distortion objectives. The detailed implementation is presented in Algorithm~\ref{alg:solution2}.

% The detailed algorithms for both solutions are presented in Supplementary Material Section~\ref{sec:detailed_alg} as Algorithm~\ref{alg:solution1} and Algorithm~\ref{alg:solution2}.

%% file: sec/4_experiment.tex
\section{Experimental results}

\subsection{Experimental settings}

\textbf{Training.}
We use the COCO 2017 dataset~\cite{lin2014microsoft} for training, which contains 118,287 images, each having around 640×420 pixels. We randomly crop 256×256 patches.

Following the settings of CompressAI~\cite{begaint2020compressai}, we set \(\lambda\) to \{18, 35, 67, 130, 250, 483\} × \(10^{-4}\). For all models, we train each model using the Adam optimizer with \(\beta_1 = 0.9\), \(\beta_2 = 0.999\), a batch size of 32, and an initial learning rate (lr) of \(1e{-4}\). We use ReduceLROnPlateau lr scheduler with a patience of 10 and a factor of 0.5. For ``Standard'' and ``Solution 1'' models, the \(\lambda = 0.0018\) models are trained for 200 epochs. For models with other \(\lambda\) values, we fine-tune the model trained with \(\lambda = 0.0018\) for an additional 150 epochs.  For ``Solution 2'' models, we fine-tune the corresponding ``Standard'' models for 50 epochs with an initial lr of 5e-5. Solution 1 hyperparameters are set as \(\beta = 0.025\) and \(\gamma = 0.001\).

\textbf{Testing.} Three widely used benchmark datasets, including Kodak~\cite{KodakWeb}, Tecnick~\cite{asuni2014testimages}, and CLIC 2022~\cite{CLIC2022}, are used to evaluate the performance of the proposed method.

\subsection{Quantitative results}
\label{subsec:qr}
We compare our proposed method with standard trained models on prevalent LICs including M\&S Hyperprior~\cite{minnen2018joint}, ELIC~\cite{he2022elic}\footnote{For the training of  ELIC models, we do not apply the mixed quantization strategy~\cite{minnen2020channel,he2022elic}; instead, we use uniform noise as the quantization method.}, and TCM-S~\cite{liu2023learned} to demonstrate its performance. We use standard rate-distortion trained models as the anchor to compute BD-Rate~\cite{BDrate}.

Table~\ref{tab:bdrate} shows the BD-Rate improvement achieved by our proposed method compared to baseline approaches across three datasets. Our method consistently surpasses all baselines on both common and high-resolution test datasets, underscoring its effectiveness. For example, ELIC Solution 1 achieves BD-Rate improvement of -2.68\%, -2.63\%, and -3.00\% compared to ELIC Standard on the Kodak, Tecnick, and CLIC2022 datasets, respectively, demonstrating the substantial improvements our approach provides. Furthermore, ELIC Solution 2, which fine-tunes existing standard R-D optimized models, achieves BD-Rate improvement of -1.81\%, -2.03\%, and -2.39\% on the three test datasets, illustrating the efficacy of our fine-tuning strategy. Overall, Solution 1 yields slightly better results than Solution 2, as Solution 2 is used only for fine-tuning over a few epochs.

\begin{figure*}[htbp] 
\newcommand{\mywidth}{0.324}
\centering 
\begin{subfigure}[b]{\mywidth\linewidth}
    \centering
    \includegraphics[width=\linewidth]{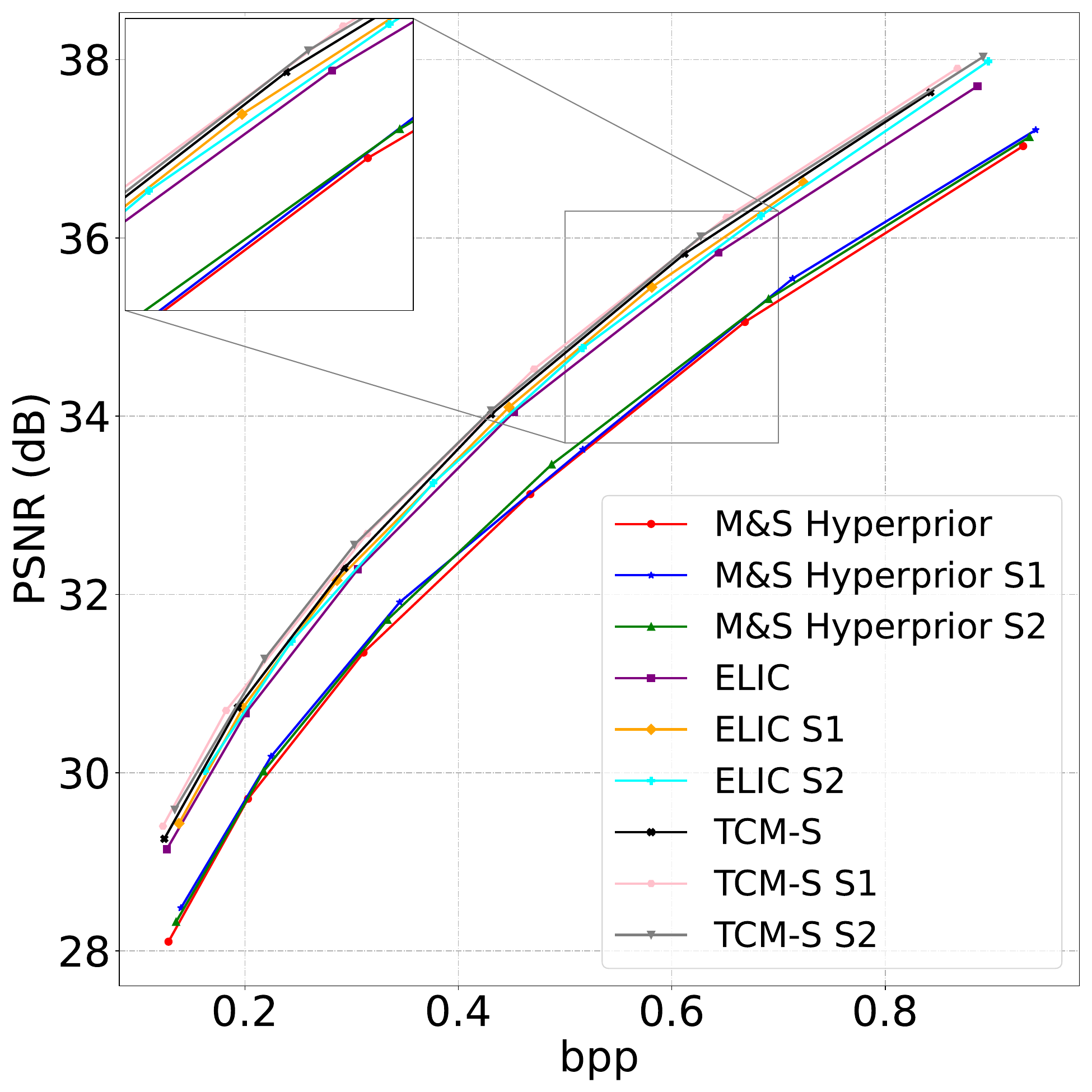}
    \caption{Kodak}
    \label{subfig:kodak}
\end{subfigure}
\hfill
\begin{subfigure}[b]{\mywidth\linewidth}
    \centering
    \includegraphics[width=\linewidth]{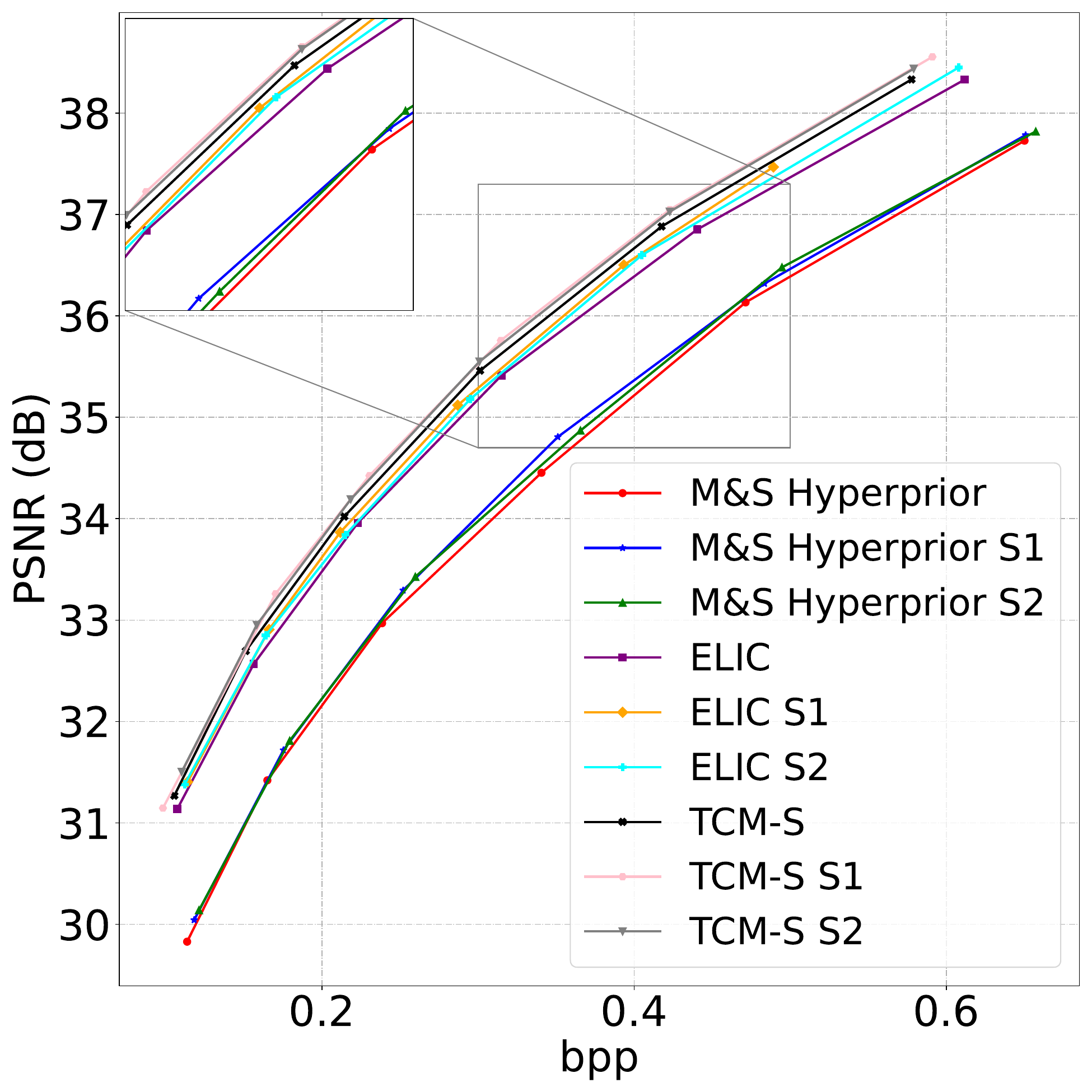}
    \caption{Tecnick 1200x1200}
    \label{subfig:Tecnick}
\end{subfigure}
\hfill
\begin{subfigure}[b]{\mywidth\linewidth}
    \centering
    \includegraphics[width=\linewidth]{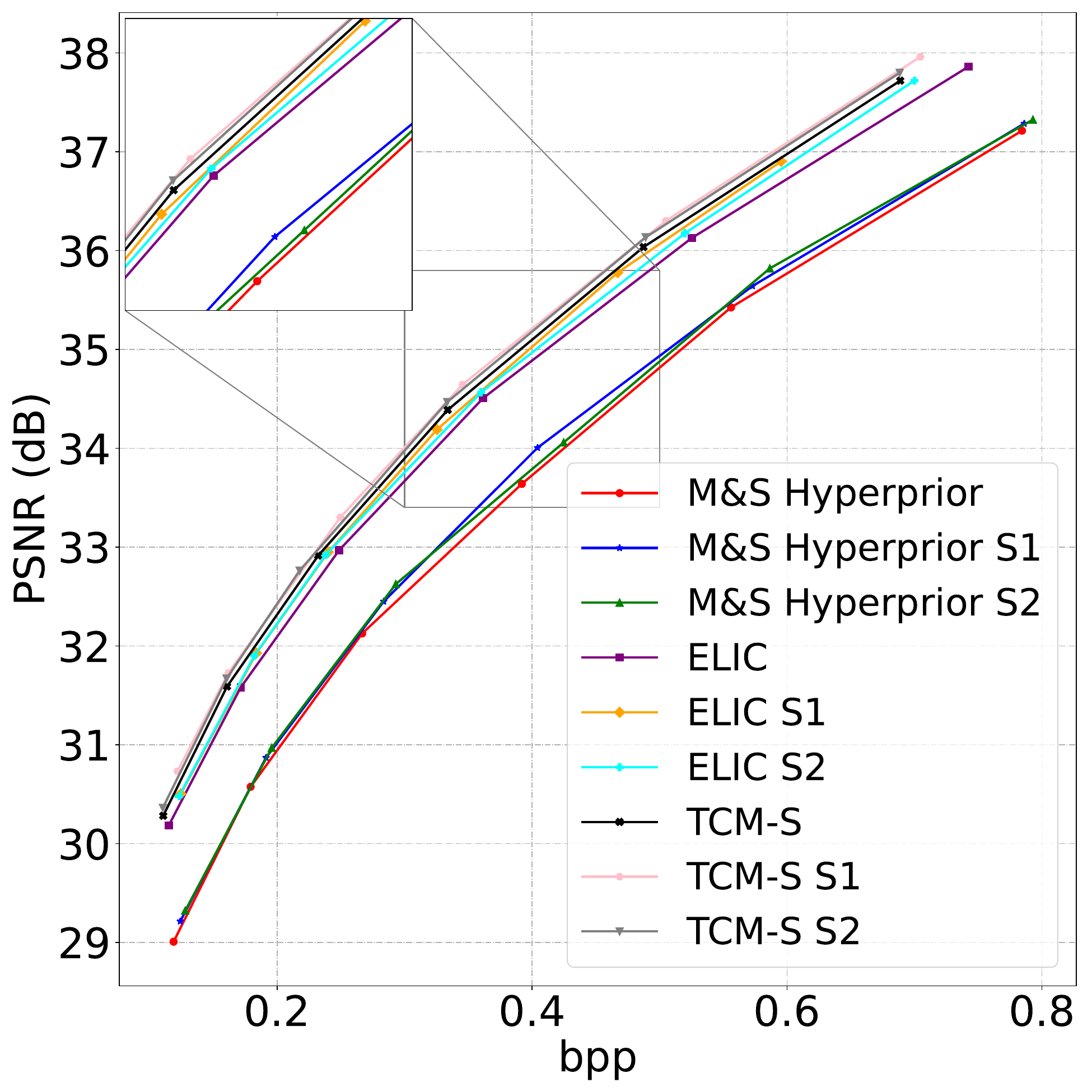}
    \caption{CLIC 2022}
    \label{subfig:CLIC}
\end{subfigure}
\caption{\textbf{R-D curves of various methods. }{\it Please zoom in for more details}.} 
\label{fig:rd_fig} 
\end{figure*}

Fig.~\ref{fig:rd_fig} further illustrates the R-D curves for all methods. The proposed method consistently outperforms the standard optimized models, particularly in the high bpp range, demonstrating its superior R-D performance

\begin{table}[htbp]
    \centering
    \caption{BD-Rate Compared to Standard R-D optimization}
      \resizebox{1\linewidth}{!}   {
    \begin{tabular}{c|c|c|c|c}
        \hline \hline
        \multicolumn{2}{c|}{\multirow{2}{*}{Method}} &  \multicolumn{3}{c}{BD-Rate (\%) $\downarrow$} \\  \cline{3-5}
        \multicolumn{2}{c|}{} &  Kodak & Tecnick & CLIC2022 \\ \hline
        \multirow{3}{*}{\makecell{M\&S\\Hyperprior~\citep{minnen2018joint}}} &  Std.  & 0\% & 0\%&0\% \\ 
        &  Sol. 1  & \textbf{-1.95\%} &\textbf{-1.91\%} &\textbf{-2.03\%} \\ 
        &  Sol. 2  &-1.64\%&-1.55\% & -1.81\%\\\hline 
        \multirow{3}{*}{ELIC~\citep{he2022elic}} &  Std.  &0\% &0\% &0\% \\ 
        &  Sol. 1  &\textbf{-2.68\%} &\textbf{-2.63\%} &\textbf{-3.00\%} \\ 
        &  Sol. 2  &-1.81\% &-2.03\% &-2.39\% \\\hline 
        \multirow{3}{*}{TCM-S~\citep{liu2023learned}} &  Std.  & 0\%& 0\%& 0\%\\ 
        &  Sol. 1  & \textbf{-2.53\%} &\textbf{-2.73\%} & \textbf{-2.66\%}\\ 
        &  Sol. 2  &  -1.87\%&-2.09\% &-2.12\% \\
        \hline \hline
    \end{tabular}}
    \begin{tablenotes}
  % \centering
  \item ``Std.'' denotes Standard, and ``Sol.'' denotes Solution. \textbf{Bold} indicates the best result.%, while \textcolor[rgb]{ 0,  .69,  .941}{blue} indicates the second best.
  \end{tablenotes}
    \label{tab:bdrate}
\end{table}

\subsection{Complexity}
\label{subsec:compl}

As shown in Table~\ref{tab:comp}, our proposed methods introduce a moderate computational overhead compared to standard R-D optimization. Solution 1 requires an additional weight update (\( w_t \)) by performing two forward passes, which increases epoch training time by approximately 20\%. Solution 2, which involves calculating the Hessian matrix through matrix multiplications and performing matrix inversions, adds an overhead of 47–50\%. For example, in the M\&S Hyperprior~\citep{minnen2018joint} model, the computational cost (10 minutes, normalized to 100\%) increases by 20\% with Solution 1 and by 50\% with Solution 2. Similar trends are observed for ELIC~\citep{he2022elic} and TCM-S~\citep{liu2023learned}. Although Solution 2 incurs a higher computational cost, it is only applied during fine-tuning over fewer epochs. Both methods demonstrate significant performance gains that justify the additional complexity. Importantly, our methods do not affect the inference phase of the trained models, leaving inference time unchanged.

\begin{table}[htbp]
    \centering
    \caption{Complexity Compared to Standard R-D optimization}
    \begin{tabular}{c|c|c}
        \hline \hline
        \multicolumn{2}{c|}{{Method}} & $\Delta$ Epoch Training Time $\downarrow$ \\  \hline
        \multirow{3}{*}{\makecell{M\&S\\Hyperprior~\citep{minnen2018joint}}} &  Std.  & 10 mins\\ %10min
        &  Sol. 1  & +20\% \\ %12min
        &  Sol. 2  & +50\% \\\hline %15min
        \multirow{3}{*}{ELIC~\citep{he2022elic}} &  Std.  & 35 mins\\ %35min
        &  Sol. 1  & +20\% \\ %42min
        &  Sol. 2  & +47\% \\\hline 
        \multirow{3}{*}{TCM-S~\citep{liu2023learned}} &  Std.  &  60 mins \\ %60min
        &  Sol. 1  & +23\% \\ %74min
        &  Sol. 2  &+48\% \\\hline %
        \hline 
    \end{tabular}
    \label{tab:comp}
\end{table}

\subsection{Ablation study}
\label{subsec:ablation}
We conduct comprehensive experiments to find the impact of various factors of the proposed method. All experiments are conducted on the M\&S Hyperprior~\cite{minnen2018joint} model with $\lambda = 0.013$. The other settings are the same as the main experiments. In the R-D plane (bpp-PSNR figures), the upper left represents better results.
\begin{figure*}[htbp] 
\newcommand{\mywidth}{0.33}
\centering 
\begin{subfigure}[b]{\mywidth\linewidth}
    \centering
    \includegraphics[width=\linewidth]{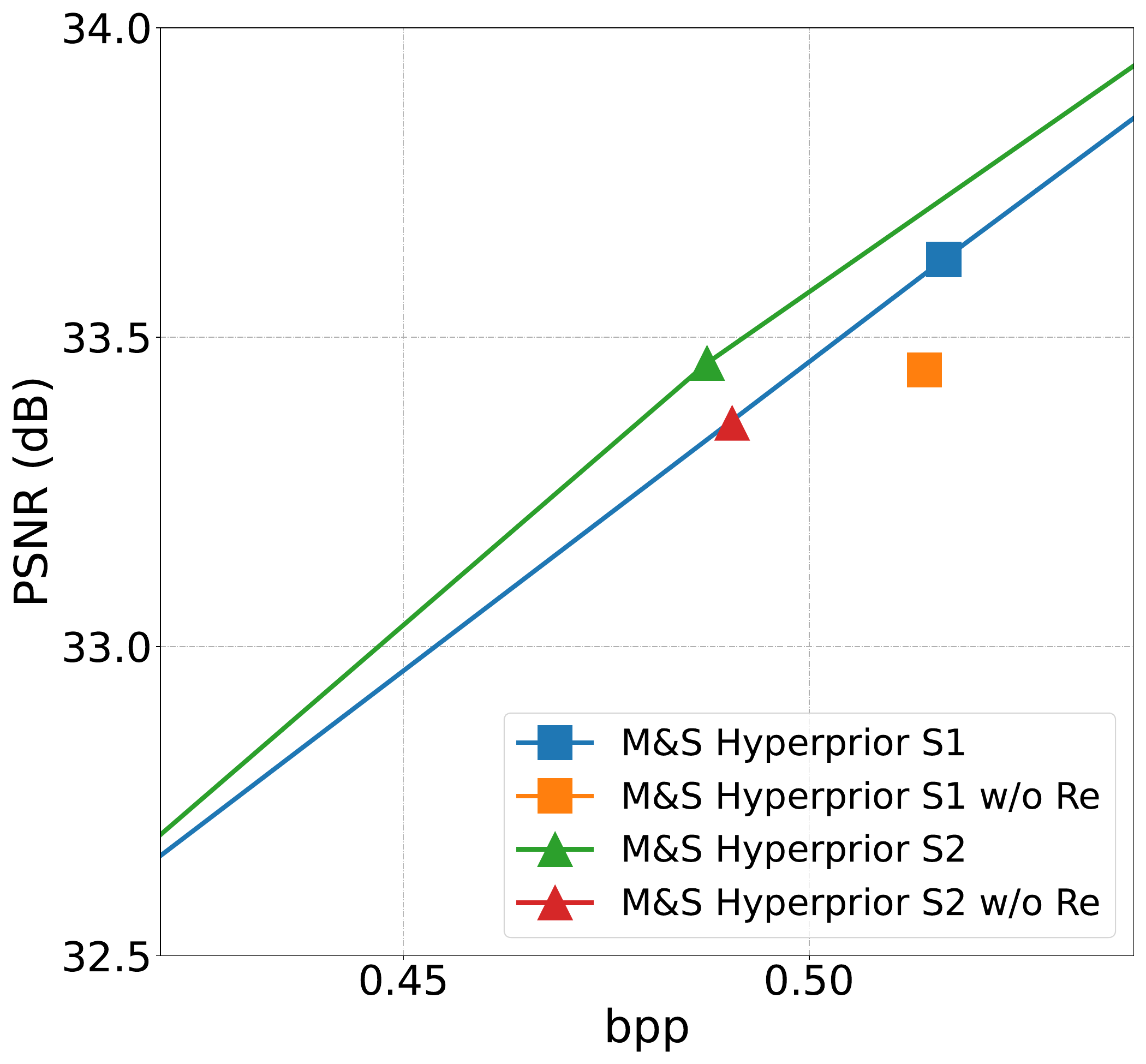}
    \caption{Renormalization}
    \label{subfig:renorm}
\end{subfigure}
\hfill
\begin{subfigure}[b]{\mywidth\linewidth}
    \centering
    \includegraphics[width=\linewidth]{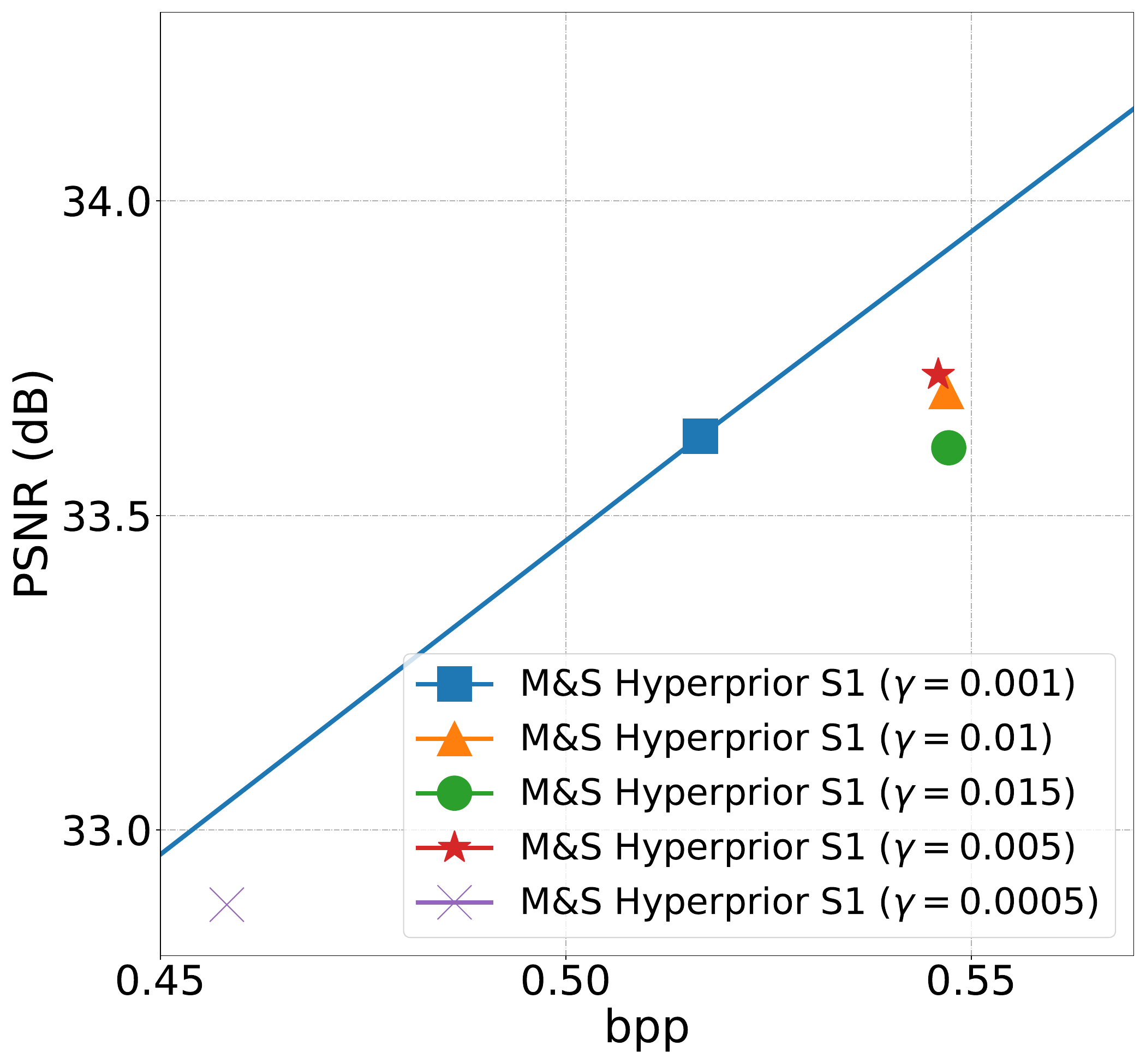}
    \caption{Weight decay}
    \label{subfig:wd}
\end{subfigure}
\hfill
\begin{subfigure}[b]{\mywidth\linewidth}
    \centering
    \includegraphics[width=\linewidth]{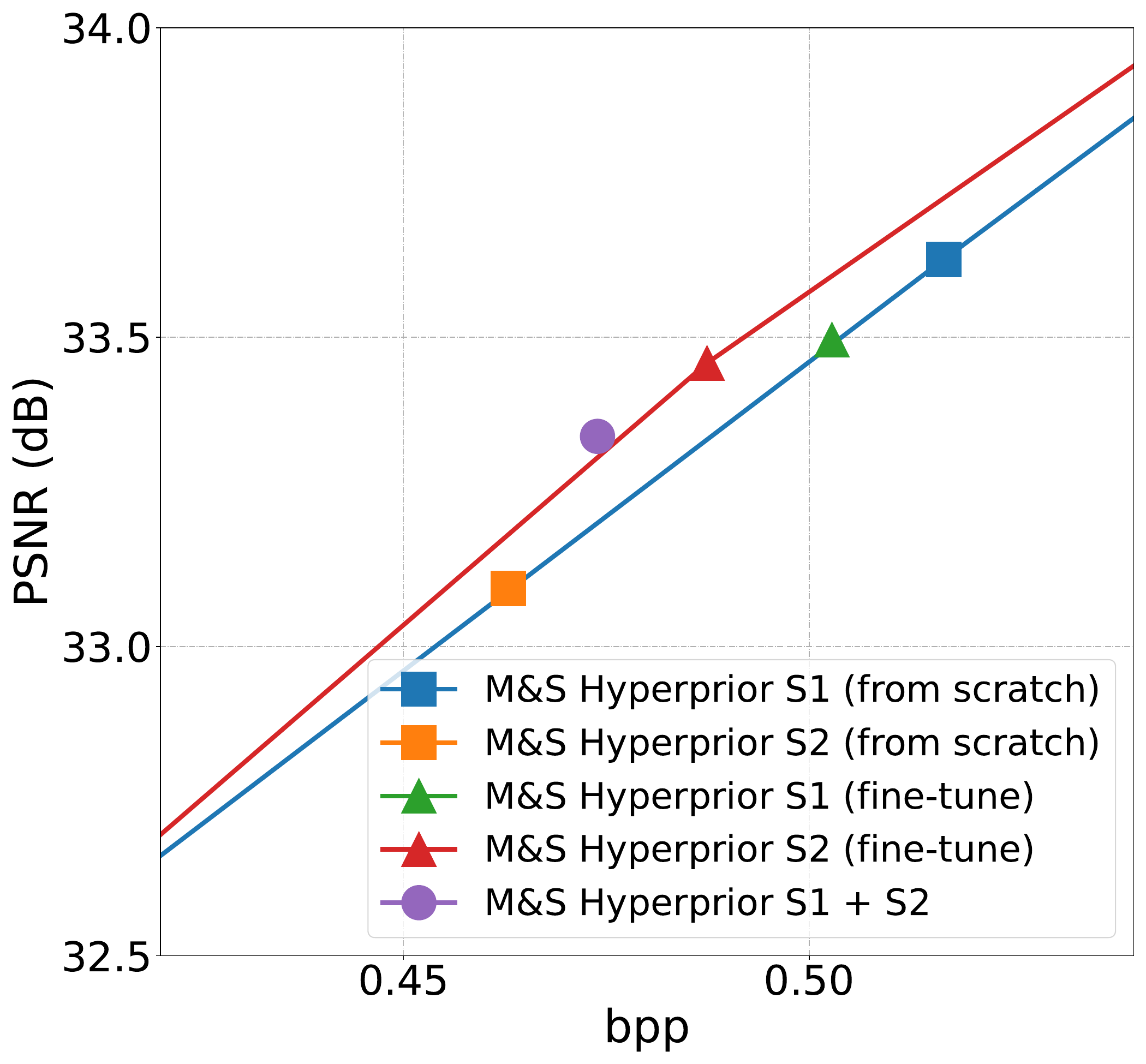}
    \caption{Cross validation}
    \label{subfig:cross}
\end{subfigure}
\caption{Ablation experiments on proposed methods. M\&S Hyperprior model, \(\lambda\) = 0.013. Kodak dataset.} 
\label{fig:ab_fig} 
\end{figure*}

\subsubsection{Weight renormalization}
In the end stages of the training process, the loss \( \mathcal{L}_{i,t} \) can become small, which may lead to numerical instability due to the multiplicative factor \( \frac{w_{i,t}}{\mathcal{L}_{i,t}} \). To address this, we apply weight renormalization to the proposed solutions. We experimented with removing this renormalization, and the results (``w/o Re''), shown in Fig.~\ref{subfig:renorm}, indicate that renormalization improves the final convergence of both solutions.

\subsubsection{Weight decay}  
In Solution 1, we introduce a weight decay term when updating the unconstrained softmax logits to mitigate potential non-convergence issues that arise from instantaneous gradients~\cite{zhou2022convergence}. To evaluate the effect of weight decay, we perform a series of experiments, varying the decay coefficient \( \gamma \) across values \( \{0.01, 0.015, 0.005, 0.001, 0.0005, 0\} \), as shown in Fig.~\ref{subfig:wd}. Notably, setting \( \gamma = 0 \) (i.e., no weight decay) results in non-convergence issues that precluded plotting the results. Among the other tested values, \( \gamma = 0.001 \) yields the best results, and we thus adopt \( \gamma = 0.001 \).

\subsubsection{Solutions cross-validation}  
As discussed, Solution 1 is a coarse-to-fine gradient descent method along standard R-D optimization trajectories that iteratively refine the gradient weights $w_{i,t}$, whereas Solution 2 is a constrained QP approach that offers an analytically precise solution. Solution 2 is more time-intensive. Intuitively, we apply Solution 1 for training models from scratch and Solution 2 for fine-tuning pre-trained models. To validate this strategy, we reverse these roles: using Solution 1 to fine-tune pre-trained models and Solution 2 to train models from scratch. The results, shown in Fig.~\ref{subfig:cross}, reveal that Solution 1 (fine-tuning) performs worse than Solution 2 (fine-tuning), while Solution 1 (from scratch) achieves comparable performance to Solution 2 (from scratch) but requires less training time. This outcome supports our hypothesis: Solution 1 is well-suited for training from scratch, where it can operate in a coarse-to-fine manner along the optimization trajectory, while Solution 2's precision and time-consuming make it more suitable for fine-tuning. Additionally, applying Solution 2 to further fine-tune Solution 1 models only yields marginal improvements compared to fine-tuning standard models, making cascaded usage of both solutions unnecessary.

%% file: sec/5_conclusion.tex
\section{Conclusion}
In this paper, we tackle the imbalance update in the R-D optimization of LICs by framing it as a MOO problem. Viewing R-D optimization as a multi-objective task, we propose two solutions to address the imbalance caused by gradient divergence. Solution 1 uses a coarse-to-fine approach along the standard R-D optimization trajectory, employing gradient descent to refine the gradient weights iteratively. This method is well-suited for training LIC models from scratch, as it progressively balances rate and distortion objectives throughout the whole training process. Solution 2 formulates the balanced R-D optimization as a constrained-QP problem, providing a precise solution by solving the KKT conditions. Although Solution 2 requires more computation, it offers an accurate adjustment of gradient weights, making it ideal for fine-tuning pre-trained models. Extensive experiments validate the effectiveness of both solutions, showing consistent R-D performance improvements over standard training schemes across various LIC models and datasets with acceptable additional training costs.

\textbf{Limitations and Future Work.} Our approach balances the R-D optimization of LICs by introducing an additional subproblem~\cite{liu2024famo,sener2018multi}, which moderately increases training complexity. Future work could focus on designing more efficient algorithms to mitigate this added computational cost. Recent research indicates that Tchebycheff Scalarization~\cite{lin2024few,lin2024smooth} is a promising avenue, offering favorable theoretical properties along with low computational complexity. Additionally, we employ the softmax function to constrain weights within the probability simplex; however, projection-based methods could be also explored~\cite{krichene2015efficient,wang2013projection}.